\documentclass[prl,twocolumn,showpacs]{revtex4}

\usepackage{graphicx}
\usepackage{bm}

\begin{document}

\def\e{\epsilon}
\def\mat#1{\bm{#1}}
\def\Q{\vec{Q}}
\def\k{\vec{k}}
\def\kp{\vec{k'}}
\def\kk{\kp}
\def\q{\vec{q}}
\def\etal{{\em et.\ al.}}   
\def\komma{\; ,\;}
\def\punkt{\; .\;}
\def\correlation#1#2{\ll \! #1 | #2 \! \gg}
\def\expect#1{\langle #1 \rangle}
\def\onepart#1#2#3{ G_{#1, #2}(#3)}
\def\as{\alpha\sigma}
\def\ket#1{|#1\rangle}
\def\bra#1{\langle  #1 |}
\def\eas{\Delta E}
\def\invmat#1{\left[ #1 \right]^{-1}}
\def\IV{intermediate valence}

\title{Composite spin and orbital triplet superconductivity formed out
of a Non-Fermi Liquid phase}

\bigskip
\author{Frithjof B. Anders}
\affiliation{Institut f\"ur Festk\"orperphysik, Technical
University Darmstadt, 64289 Darmstadt, Germany}
\date{\today}

\begin{abstract}
An unconventional superconducting phase is explored
developing  out of a non Fermi-liquid phase of the two-channel
Anderson lattice model. It is characterized  by a composite order parameter
comprising of a local spin or orbital degree of freedom bound to 
triplet Cooper pairs with an isotropic and a nearest neighbour form
factor. 
The superconducting transition temperature peaks in the moderate
intermediate valence regime and 
vanishes at integral valence. The gap function is non
analytic and odd in frequency, and a pseudo-gap develops in the
conduction electron density of states which vanishes as $|\omega|$
close to $\omega=0$. 
\end{abstract}

\maketitle

\paragraph*{Introduction.}  

Heavy Fermion (HF) superconductivity \cite{Grewe91} has drawn
much attention since the discovery of
superconductivity in CeCu$_2$Si$_2$ \cite{Steglich79} which is likely
characterized by an anisotropic order parameter with  symmetry yet  to be
determined. It became apparent over the last decade that almost all
HF materials are unstable  with respect to magnetic or superconducting
phase transitions, which either compete with each other or can even
coexist as found in uranium based materials. Non-Fermi liquid behaviour
is often found in the vicinity of a so-called quantum critical point 
which is a point in parameter space where the ordering temperature is
quenched to $T=0$ by a control parameter like pressure or
doping \cite{Hertz76}. Doped CeCu$_{6-x}$Ag$_x$ is one of the most prominent
examples \cite{Loehneysen96}.

Two channel Anderson and Kondo lattice models, however, 
also exhibit non-Fermi liquid
behaviour in the paramagnetic phase driven by unquenched and
fluctuating local degrees of freedom \cite{AndersJarCox97,JarrellPangCox97}.
This phase is characterized by a large residual resistivity and
entropy, and ill defined electronic quasi-particles. Fermi liquid
physics is restored by cooperative ordering or applied magnetic field
or stress \cite{JarrellPangCox97,Anders99}.
In particular, for a non Kramers doublet crystal field state in U$^{4+}$
or Pr$^{3+}$ ions, a two channel Kondo effect is possible
\cite{Cox87}. The magnetic ``spin'' of the electrons serves as channel
index in this case. 
UBe$_{13}$ and PrFeP$_{12}$ are prominent candidates for a quadrupolar
Kondo lattice description, as their enormous resistivity
($>100\mu\Omega cm$) is removed only by phase transitions
(superconductivity in UBe$_{13}$, antiferroquadrupolar order in
PrFeP$_{12}$). Indeed, it has been shown that commensurate and
incommensurate orbital ordering, as well as ferromagnetism are
possible in this model depending upon coupling strength and
filling \cite{Anders99}. Evidence of a first order transition to an odd
frequency  pairing state in the Kondo limit (near integral valence) 
has been adduced, 
which  is a singlet in both spin and
channel indices, and no $\q$ value was preferred for the center of mass
momentum (COMM) \cite{JarrellPangCox97}.

In this paper, the possibility and nature of a superconducting phase
transition out of the NFL paramagnetic phase within dynamical mean
field theory (DMFT) \cite{Pruschke95,Georges96}  will be explored. We
show that in a intermediate valence (IV), two channel Anderson
lattice  a 2nd order transition to a triplet spin, triplet channel
spin (StCt) order parameter with zero COMM will develop. The gap
function is odd in frequency and singular, leading to a quasi-particle
density of states that vanishes linearly with energy, in agreement
with specific heat and spin lattice relaxation data for many HF
materials. The order may arise form either two channel quadrupolar or
magnetic ground states, and the transition temperature peaks in the
moderate IV regime and vanishes at integral valence. The order
parameter  can be viewed as a composite of a bound spin and an even
frequency triplet Cooper pair.  Moreover, our analysis shows
that isotropic  and   spatially extended states
contribute to the superconducting phase, where
the latter dominate in the IV regime.
The binding of the fluctuating local degrees of freedom
to the itinerant Cooper pair  in the superconducting state is a natural
consequence of the tendency to remove their residual entropy similar
to a magnetic phase transition, a concept which was  introduced first
by Abrahams \etal\ \cite{Abrahams95}.

\paragraph*{Composite order parameter.} 
The two-channel periodic Anderson model   \cite{AndersJarCox97} 
 \begin{eqnarray}
\label{eq:tca-97}
\label{equ-hamil}
  \hat H  &= &
\sum_{\k\alpha\sigma} \e_{\k}
c^\dagger_{\k\alpha\sigma}c_{\k\alpha\sigma}
+\sum_{i\sigma} E_\sigma X_{\sigma,\sigma}^{(i)}
+\sum_{i\alpha} E_\alpha X_{\alpha,\alpha}^{(i)}
\nonumber\\  &   &
 +
\sum_{i\sigma\alpha} (-1)^{1+\alpha}V\left\{
c^\dagger_{i\alpha\sigma}  X_{-\alpha,\sigma}^{(i)}
+ h.c
 \right\} 
\end{eqnarray}
describes the coupling of two degenerate conduction bands,  labelled by
a spin $\sigma$ and a channel index $\alpha=\pm1$,
to two localized doublets at each lattice site  via hybridization
matrix elements $(-1)^{1+\alpha}V$. $X$ are 
 Hubbard operators, and $\e_{\k\as}$ the band dispersion.
All energies will be given in units of $\Delta = V^2\pi{\cal N}_f$, the
hybridization width.
Since the model (\ref{equ-hamil})  conserves spin and orbital quantum
numbers, the symmetry of the conduction electron pair operator is
classified in four sectors, comprising of the product space of spin
and channel singlets or triplets.
Introducing the transposed bi-spinor
$\vec{\psi}^T(\k) =  \left( c_{\k+\uparrow},c_{\k+\downarrow},
c_{\k-\uparrow},c_{\k-\downarrow}\right)$, the matrix elements of the
tensor pair operator in the 
triplet/triplet (StCt) sector 
are given by
\begin{equation}
  \label{eq:pair-t-t}
  P^{t,t}_{ij}  = \frac{1}{N} \sum_{\k} S(\k)
\vec{\psi}^T(\k)i\mat{\sigma}_{y} \, \mat{\sigma}_i
 \;i\mat{\tau}_y \, \mat{\tau}_j \vec{\psi}(-\k) 
\komma
\end{equation}
where $\mat{\vec{\sigma}}$ acts in the spin sector and
$\mat{\vec{\tau}}$ in the channel sector. 
 The local nature of the DMFT permits  only the Cooper
pairs whose $\k$-dependent form factor $S(\k)$ transforms according to
$\Gamma_1$, the trivial irreducible representation. 
The pair expectation values  $\expect{P^{t,t}_{ij}}$, however,
always must vanish due to Pauli's principle.
As a consequence, the anomalous Green functions in this sector have
to be odd in frequency. This peculiar kind of pairing was  first 
suggested for He$^{3}$ \cite{Berezinskii74} and
later revitalized  \cite{BalatskyAbrahams92} in the context
of {\em High-T$_c$}. Coleman \etal\
\cite{ColemanMirandaTsvelik94} investigated the possibility of an
odd-frequency pairing for a single channel Kondo lattice model and found
a staggered order parameter with finite COMM $\Q$.  However, their
mean-field solution breaks 
spin-rotational invariance. 

Abrahams \etal\ noticed that the time derivative $C_{\gamma\gamma'}=
\left.\expect{\frac{d}{d\tau}c_\gamma(\tau)
c_{\gamma'}}\right|_{\tau=0}$  is non-vanishing in the odd frequency
superconducting phase and may possibly furnish an order parameter 
\cite{Abrahams95}. The derivative $\frac{d}{d\tau}c_\gamma(\tau)$ is
equivalent to the commutator $[H,c_\gamma(\tau) ]$ and consists of a product of
two operators in the case of the two-channel periodic Anderson model
(\ref{equ-hamil}): a conduction electron pair operator and a local
magnetic   or channel spin operator.
Thus, one arrives at a composite order parameter, which correlates
local and itinerant degrees of freedom: formation of band Cooper
pairs is stimulated in the presence of a local pair resonating between
definite spin and channel states with proper symmetry.
Using  equation of motion  techniques on 
(\ref{eq:tca-97}), we derive with the 
definitions
\begin{equation}
D_{\sigma\sigma'}^{\alpha\alpha'}  =
\frac{1}{\beta}\sum_{i\omega_n} e^{i\omega_n\delta} 
i\omega_n\frac{1}{N}\sum_{\k}
\onepart{c_{\k\alpha\sigma}}{c_{-\k\alpha'\sigma'}}{i\omega_n}
  \label{eq:isotropic-wave}
\end{equation}
and
- $\onepart{c_{\k\alpha\sigma}}{c_{-\k\alpha'\sigma'}}{i\omega_n}$ is the
Fourier transform of the anomalous Green function
$- \langle
T (c_{\k\alpha\sigma}(\tau) c_{-\k\alpha'\sigma'}\rangle$ -
\begin{equation}
   T_{\sigma\sigma'}^{\alpha\alpha'} 
 = 
 \frac{1}{\beta}\sum_{i\omega_n} e^{i\omega_n\delta} i\omega_n
\frac{1}{N}\sum_{\k} \e_{\k\as}\onepart{c_{\k\as}}{c_{-\k\alpha'
\sigma'}}{i\omega_n}  
  \label{eq:extended-s-wave}
\end{equation}
the {\em exact relations} for the symmetrized
order parameter  components for the StCt sector 
\begin{eqnarray}
 O_{ij} &= & 
\mbox{sign}(\Delta E)\left[ \expect{s_i P^{s,t}_{j}}
- \expect{\tau_j P^{t,s}_{i}}\right]
\label{eqn:stct-order-param}
\\
 &= & 
\left(|\Delta E|D_{ij}
+\mbox{sign}(\Delta E)
T_{ij}
\right)/V^2
\komma
\label{equ-tt-order-parameter}
\end{eqnarray}
where $i,j=x,y,z$, $i(j)$ indexes the spin(channel) space,
$  \vec{P}^{s,t}  = 
\frac{1}{N} \sum_{\k} S(\k) \vec{\psi}^T(\k)i\mat{\sigma}_y
 i\mat{\tau}_{y}\mat{\vec{\tau}}  \vec{\psi}(-\k) 
$ is the SsCt, and
$\vec{P}^{t,s}  = \frac{1}{N} \sum_{\k} S(\k)
  \vec{\psi}^T(\k)i\mat{\tau}_y
 i\mat{\sigma}_{y}\mat{\vec{\sigma}}\; \vec{\psi}(-\k) 
$ the StCs pair operator.  Applying (\ref{eq:pair-t-t}), the symmetrized
 pair expectation values $D_{ij}$ ($T_{ij}$) are given by
$ D_{ij} (T_{ij}) = \sum_{\sigma\sigma´}^{\alpha\alpha'}
D_{\sigma\sigma'}^{\alpha\alpha'}  (T_{\sigma\sigma'}^{\alpha\alpha'})
[i\mat{\sigma}_{y} \,
\mat{\sigma}_i]_{\sigma\sigma'}
[i\mat{\tau}_y \, \mat{\tau}_j ]_{\alpha\alpha'}
$.
Isotropic 
and  modulated Cooper pairs  with a form factor
$S(\k)= \e_{\k}$ contribute differently to the local composite order
parameter  $O_{ij}$ in different regimes: in the
IV regime, i.\ e.\ $|\Delta E = E_\sigma-E_\alpha|/\Delta \le 1$,
modulated pairs dominate which is consistent with strong hybridization
between $f$-shell electrons and conduction electrons,
the isotropic pair contribution vanishing at the degeneracy point $\Delta E =0$.
The composite order parameter $O_{ij}$ (\ref{eqn:stct-order-param}) is
invariant under exchange of the local doublets, and therefore
superconductivity 
can arise from either a magnetic or quadrupolar ground state.

\begin{figure}[t]
  \begin{center}
   \begin{tabular}{cc}
\includegraphics[width=55mm]{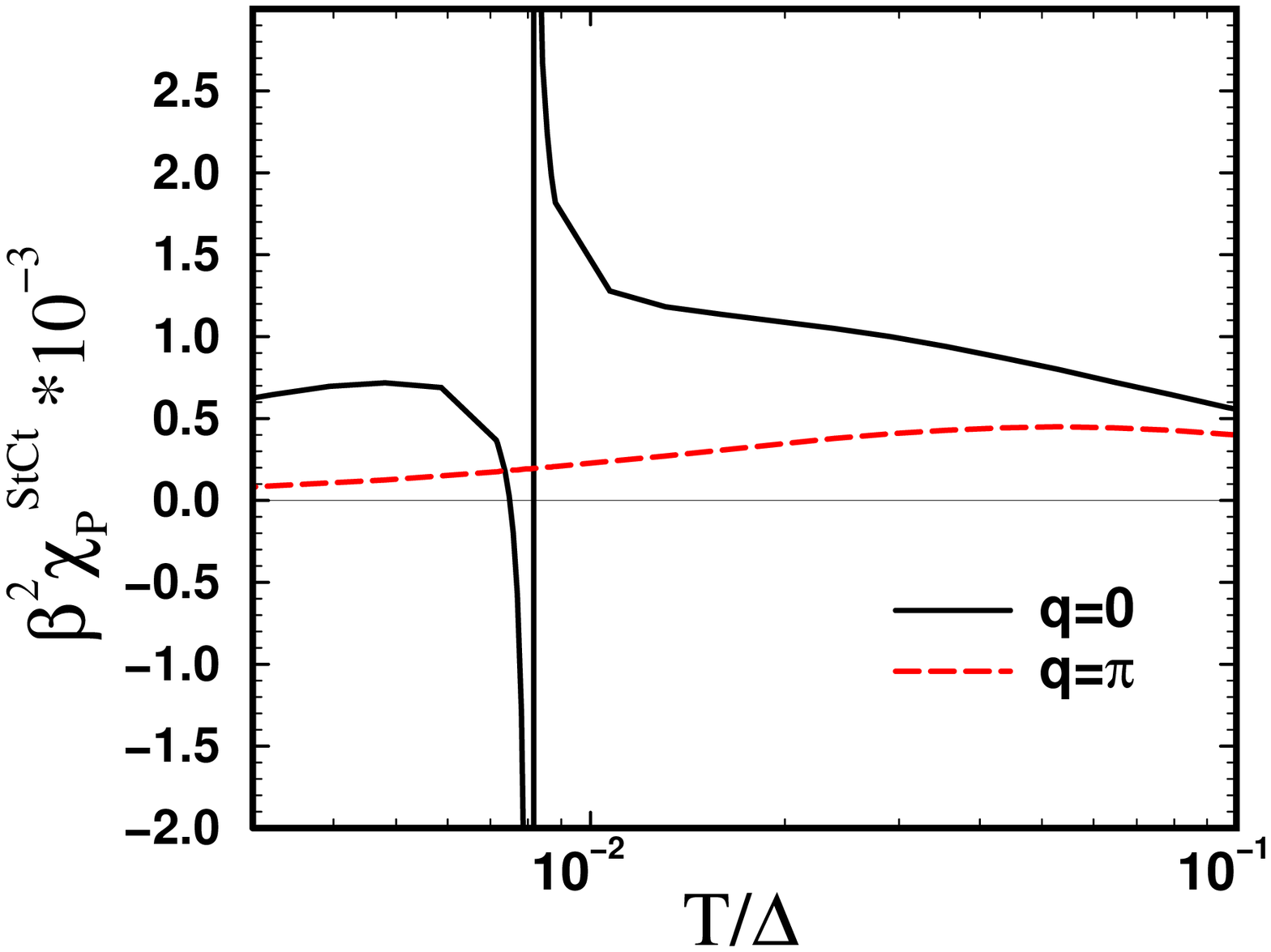}
&
\includegraphics[width=24.5mm]{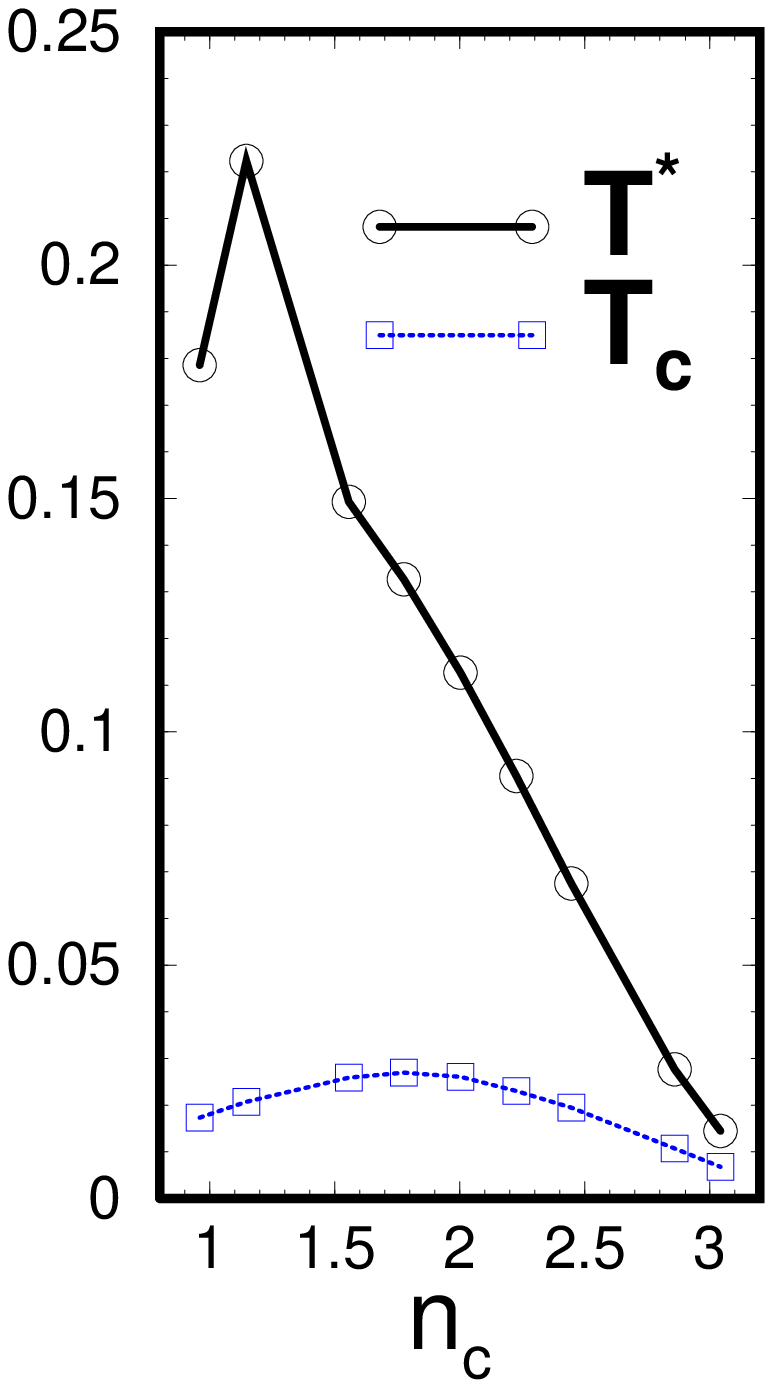}
\\
(a) & (b)
 \end{tabular}
\caption{(a) Pair susceptibility $\chi_{P}^{StCt}$ vs temperature in
  the triplet/triplet (StCt)  sector for $q=\pi$ (staggered) and
  $\q=0$.   Parameters: $\eas = -3$ and $n_c =2$. (b) Dependence of
  $T_c$ and characteristic temperature $T^*$ on  $n_c$, for $\eas =
  -\Delta$. }
\label{fig:pair-sus-CtSt}
\end{center}
\end{figure}

\paragraph*{Pair Susceptibility.}

An extensive discussion of the physical nature of the possible order
parameters in the StCt sectors is motivated by the
observation, that the channel and spin exchanging local
particle-particle (pp) $f$-Green function $\Pi^{ex}_{loc}$ is purely
odd in the  incoming and outgoing frequencies for $T\rightarrow 0$, in  addition to
an $1/|w|$ singularity \cite{CoxZawa98} favouring an
odd-frequency gap function. Eqn.\ (\ref{equ-tt-order-parameter}) connects
the composite order parameter operator to derivatives of the pair
operators in frequency space  $\dot P^{odd}_{\gamma\gamma'}(i\omega_n)=
i\omega_n c_\gamma(i\omega_n)c_{\gamma'}(-i\omega_n)$
($\gamma=\k\as$). Thus, the pair susceptibility 
contains the same information as the composite order parameter susceptibility.
The  irreducible pp interaction 
$\Gamma^{tt}$ was obtained as a matrix in Matsubara frequency space  
\begin{equation}
  \label{eq:eff-pp-vertex-tt}
  \frac{1}{\beta} \mat{\Gamma}^{tt} =
\left[\mat{\chi}^{pp}_{loc}\right]^{-1}
-
\beta
\left[\mat{\Pi}^{dir}_{loc}+ \mat{\Pi}^{ex}_{loc}\right]^{-1}
\punkt
\end{equation}
 by well established procedures within the  DMFT 
\cite{JarrellPangCox97,Anders99,Pruschke95,Georges96}.
The local two particle band GF matrix $\mat{\chi}^{pp}_{loc} =\sum_{\q}
\mat{\chi}^{pp}(\q)/N$ is given by
$\chi^{pp}_{loc}(i\omega_n,i\omega_m) =
\delta_{n,m}G_c(i\omega_n)G_c(-i\omega_n)$, where
$G_c$ denotes the local conduction electron Green function.
and $\Pi^{dir}_{loc}$ is the local direct pp GF. Then, the pair
susceptibility reads 
\begin{equation}
  \label{eq:pair-sus-lambda}
   \chi_{P}(\q ) = \frac{1}{\beta} |c_{\lambda_m}|^2 \frac{1}{1-\lambda_m} 
+ \mbox{regular terms}
\end{equation}
where $\lambda_m$ is the largest eigenvalue of the matrix $  \mat{M} =
\sqrt{\mat{\chi}^{pp}(\q)} \frac{1}{\beta}\mat{\Gamma}^{tt}(0)
\sqrt{\mat{\chi}^{pp}(\q)}$ in the odd frequency sector.
$c_{\lambda_m} = \sum_n i\omega_n \sqrt{\chi(\q,0)(i\omega_n)} [\ket{\lambda}]_n$
This DMFT-analog to the Eliashberg equation was first used by Jarrell
and co-workers \cite{JarrellPangCox97,Jarrell95}.

\paragraph*{StCt Sector.}

The pair susceptibilities of all nine components of the tensor order
parameter (\ref{equ-tt-order-parameter})  in the  StCt sector are
equal, as expected on symmetry grounds. No superconducting instability
was previously found for the two-channel Kondo-lattice model
\cite{JarrellPangCox97}  since the model is restricted to integer
$n_f$ valence. In the StCt however, $T_c$ is the largest for the IV
regime, hence charge fluctuation driven rather the spin exchange
induced. In Fig.~\ref{fig:pair-sus-CtSt}a, the pair
susceptibility $\chi_{P}^{StCt}$ is shown in the stable moment regime
for half-filling and for  COMM  $|\q|=0$  and $|\q|=\pi$.
A sign change is found at $T\approx 8\cdot 10^{-3} \Delta$  
which is an indicator for a  second order phase transition in the
triplet/triplet sector with an uniform order parameter. 

We  investigated the pair susceptibility $\chi_{P}^{StCt}$ in the IV, the
Kondo and the stable moment regime for band-fillings 
between quarter and $3/4$-filling and COMM of $|q|=0$ and
$|\q|=\pi$. A superconducting  transition  was always obtained, but only
for $q=0$. $T_c$ is filling dependent for fixed coupling constant
$g={\cal N}_f J=\Delta/|\Delta E |$ as depicted 
for the IV regime in
Fig.~\ref{fig:pair-sus-CtSt}b. $T_c$ is also compared to the
characteristic  
temperature of the lattice, $T^*$ as defined via a reduction of  the effective
moment $\mu_{eff}^2(T)=T\chi(T)$ to $0.4$ of its high
temperature value. Hence, $T^*$ serves as a phenomenological Kondo
lattice temperature (see also \cite{Anders99}).  The pair
susceptibility in the singlet/singlet sector (SsCs) agrees with the
one reported for the Kondo lattice model \cite{JarrellPangCox97}.  The
instability, however, occurs always at lower temperatures compared to
the one in the StCt sector.

\begin{figure}[t]
  \begin{center}
\includegraphics[width=85mm]{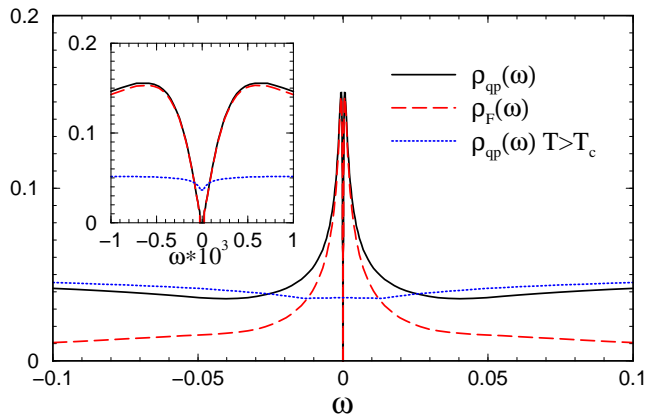}
\caption{Spectral function of the quasi-particle ($\rho_e(\omega)$)
  and of the anomalous    Green function  $\rho_F(\omega)$,   close to
  $T_c$ for $\eas =-2$ and half filling in comparison with the
  spectral function slightly above $T_c$ (dotted line). The inset
  shows a blow up of    the frequency interval $|\omega| < 10^{-3}$ to
  demonstrate the $\rho(\omega)   \propto |\omega|$ behaviour close to
  $\omega=0$. 
}
\label{fig:odd-CtSt-super-gf}
\end{center}
\end{figure}

\paragraph*{DMFT for the Superconducting Phase.}
The Nambu Green function is an $8\times 8$ matrix in  spin and
orbital space. Assuming no directional coupling between spin and
orbital degrees of freedom, the anomalous  $4\times 4$ self-energy
matrix may be written as  
$g(i\omega_n,\k) \mat{\sigma}_2\mat{\tau}_2
\left[\vec{n_s}\vec{\mat{\sigma}}\vec{n_c}\vec{\mat{\tau}} \right]$
similar to He$^{3}$ \cite{WoelfleVollhardt},  where $\vec{n_s}$ and
$\vec{n_c}$ are constant  unity vectors in spin and channel  space ,
and the  amplitude function  $g(i\omega_n,\k) $ 
has to be odd in frequency. This  reduces the full problem to a standard
$2\times 2$  size for $g(z,\k)$ and the diagonal self-energy $\Sigma$.
In order to derive  DMFT equations with a purely local
self-energy matrix $\mat{\Sigma}_c(z)$, the anomalous
self-energy is restricted to isotropic pairs, e.\ g.\ $g(z,\k) =g(z)$.
$\mat{\tilde G}_c(z)$ denotes the  medium matrix in which the
effective impurity is embedded. It is related to a generalized 
Anderson width matrix through
$  \mat{\Delta}(z) = V^2\mat{\sigma}_2 \,\mat{\tilde
    G}_c(z) \,\mat{\sigma}_2 
$
and has normal and anomalous components describing quasi-particle
propagation and pair-creation and annihilation, respectively. The
band electron self-energy $
\mat{\Sigma}_c(z) = \mat{T} \left[\mat{1} + \mat{\tilde G} \mat{T} \right]^{-1}
$
is determined by the local $T$-matrix $ \mat{T}$. The
diagonal elements of the $T$-matrix  are given by the local
quasi-particle scattering matrices $\mat{T}_{11}(z)= V^2 G_f(z)$,
and ist anomalous contribution 
is calculated from
\begin{equation}
  \label{eq:anonalous-t-matrix}
  \mat{T}_{12}(i\omega_n) = - V^4 \frac{1}{\beta} \sum_{i\omega_m}
  \Pi^{pp}_f(i\omega_n,i\omega_m;0) \tilde f (i\omega_m)
\punkt
\end{equation}
The negative sign takes into account that $\mat{\Delta}_{12}(z) = -V^2
\tilde f(z)$ where $\tilde f(z) =[\mat{\tilde G}_c(z)]_{12}$ and
$\tilde e(z) =[\mat{\tilde G}_c(z)]_{11}$. 
We obtained a solution for the superconducting phase close to $T_c$,
using the normal state 
  self-energy $\Sigma_e(z)= T_e(z)/(1+\tilde
e(z)T_e(z))$, iterating only the anomalous $T$-matrix $T_s$ and medium
$\tilde f$.
The spectra of the quasi-particle and the anomalous Green functions
are shown in Fig.~\ref{fig:odd-CtSt-super-gf}. 
Since the 
the quasi-particle (qp) scattering rate   $\Gamma_{qp}=\Im m
\Sigma(-i\delta)$ tends to become strongly reduced close to  $\omega=0$ as
expected, the full  DMFT(NCA)  turns out to be unstable in
the SC phase which is related to the well known convergence problems
in the Fermi liquid regime \cite{Kim87}.
The resulting qp density of states $\rho_{qp}(\omega)$ below $T_c$ is 
proportional to $|\omega|$ for small $\omega$ as depicted in the inset
of  Fig.~\ref{fig:odd-CtSt-super-gf}, as
can also be analytically shown using a temperature broadened
Lorentzian as an {\em ansatz} for $g(z)$  which reflects the $1/z$
singularity of $\Pi_f^{pp}$ \cite{CoxZawa98}.

\paragraph*{Discussion and Conclusion.}
There is a controversial debate (see \cite{Heid95} and reference
therein) as to whether an odd-frequency solution
with a COMM $Q=0$ is connected to a minimum or a maximum of the
free energy for a second order phase transition and, hence, may be
thermodynamically unstable.  Assuming a Fermi-liquid normal phase and
an infinitesimal odd-frequency anomalous self-energy $g(z)$ at $T_c$, Heid  
showed  that  one  indeed obtains an increase of the free energy
in the superconducting phase for $Q=0$ \cite{Heid95}. 
This also holds for a non-Fermi liquid normal phase. 

Based on the self-consistent solution for the anomalous Green function,
$g(z)$ converges to a finite value such that always a gap is generated in the
lattice  Green functions. $g(z)$ might
be approximated in leading order by a Lorentzian with a finite weight
$A=\sqrt{\Gamma_{qp}\alpha T}$. The assumption of an
infinitesimal value of $g(z)$ for all frequencies is violated and
hence Heid's theorem is not applicable in our case.
Additionally, in the presence of an anomalous medium, the quasi-particle
self-energy is also modified and, therefore, the feedback onto the effective
site must  be calculated self-consistently. We expect that the 
self-consistent  solution for the one-particle Green function in the
superconducting phase exhibits $\lim_{\omega\rightarrow 0}
\Gamma_{qp}\rightarrow 0$. In this case, a non-analytic 
$g(z) \propto A/z$, even with an small $A$,
will be able to always produce a gap. Even though  the DMFT(NCA)
equations for the superconducting  phase are
numerically unstable, they indeed indicate  a  reduction of the local free
energy in the presence of a finite anomalous Green function and a tendency
to reduce the scattering rate for the quasi-particles. 
Therefore, even though we cannot derive conclusively the type of the
phase transition, we 
believe that theses arguments favour a  second order
transition in the StCt sector. This phase transition is  associated
with a minimum of the free energy since an energy gap  develops in the
spectrum of the local band  Green function indicating an energy gain
by condensation. 

\begin{figure}[ht]
\includegraphics[width=80mm]{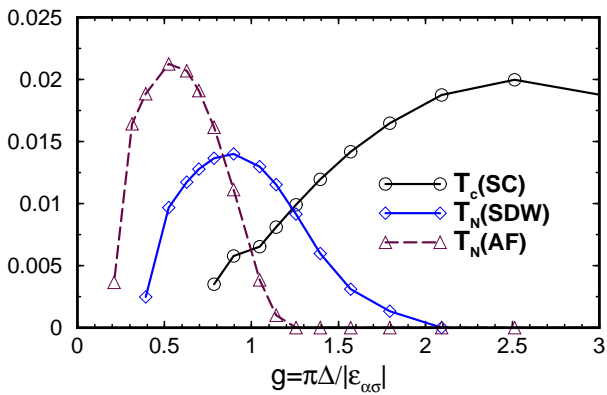}
\caption{The phase diagram of the two-channel periodic Anderson model:
    the superconducting transition temperature $T_c$, the
  antiferromagnetic  transition temperature $T_m(AF)$, the
  spin-density wave transition  temperature $T_m(SDW)$  are plotted
  versus the effective dimensionless   coupling constant
  $g=\pi\Delta/|\eas|$. 
   StCt   superconductivity dominates the IV regime  and the cross-over
  region to the Kondo regime. 
}
\label{fig:phase-diagram-magnetic-sc}

\end{figure}

The phase diagram shown for $n_c=2.2$ 
in Fig.~\ref{fig:phase-diagram-magnetic-sc}
summarizes our investigation of superconductivity and
magnetism in the two-channel Anderson model. Superconductivity
dominates the IV regime and the corresponding order parameter has spin
and channel spin triplet 
symmetry. An instability in the SsCs reported in the two-channel Kondo lattice
could be reproduced but occurs at much lower temperatures.
Spin-density wave phase
transitions take over in the Kondo-regime at $g\approx 1.3$
corresponding to $\eas=-2.4\Delta$. The SDW wave-vector is continuously shifted
towards nearest-neighbour antiferromagnetism, which is suppressed for
$g\rightarrow 0$. Since all calculations were performed in the
paramagnetic phase of the model, the highest transition temperature
defines the nature of the incipient order. One cannot rule out, however,
that an SDW or orbital ordering phase is replaced by, or even coexists with, a
superconducting phase as found in some Uranium based HF compounds.
We also would expect that the antiferroquadrupolar ordered
PrFe$_{4}$P$_{12}$ might become superconducting when  pressure is applied.

\paragraph*{Acknowledgment.}
We would like to thank D.~Cox and N.~Grewe for many stimulating
discussions. This 
work was funded in parts by an DFG grant AN 275/2-1.

\bibliography{references}

\end{document}